\documentclass[%
 reprint,
 amsmath,amssymb,
 aps,
]{revtex4-2}

\usepackage{graphicx}
\usepackage{dcolumn}
\usepackage{bm}

\usepackage{algorithmic}
\usepackage{algorithm}
\usepackage[pdfauthor={derajan}, pdftitle={How to do this}, pdfstartview=XYZ, bookmarks=true, colorlinks=true, linkcolor=blue, urlcolor=blue, citecolor=blue, pdftex, bookmarks=true, linktocpage=true，hyperindex=true ]{hyperref}
\usepackage{color}

\usepackage[inkscapelatex=false]{svg}
\begin{document}

\preprint{APS/123-QED}

\title{DAPO-QAOA: An algorithm for solving combinatorial optimization problems by dynamically constructing phase operators}
\thanks{A footnote to the article title}%

\author{Yukun Wang}
\email{yukun06@gmail.com}
\author{ZeYang Li}%
\email{lizeyang597@163.com}

\affiliation{%
 Beijing Key Laboratory of Petroleum Data Mining, China University of Petroleum, Beijing 102249,\\
 State Key Lab of Processors, Institute of Computing Technology, CAS, Beijing 100190, China
}%
\author{Linchun Wan}
\email{linchunwan@outlook.com}
\affiliation{School of Computer and Information Science, Southwest University, Chongqing 400715, China}



\date{\today}

\begin{abstract}

{The Quantum Approximate Optimization Algorithm (QAOA) is a well-known hybrid quantum-classical algorithm for combinatorial optimization problems. Improving QAOA involves enhancing its approximation ratio while addressing practical constraints of Noisy Intermediate Scale Quantum (NISQ) devices, such as minimizing the number of two-qubit gates and reducing circuit depth. Although existing research has optimized designs for phase and mixer operators to improve performance, challenges remain, particularly concerning the excessive use of two-qubit gates in the construction of phase operators. To address these issues, we introduce a Dynamic Adaptive Phase Operator (DAPO) algorithm, which dynamically constructs phase operators based on the output of previous layers and neighborhood search approach, optimizing the problem Hamiltonian more efficiently.  By using solutions generated by QAOA itself to simplify the problem Hamiltonian at each layer, the algorithm captures the problem’s structural properties more effectively, progressively steering the solution closer to the optimal target.  
Experimental results on MaxCut and NAE3SAT problems show that DAPO achieves higher approximation ratios and significantly reduces two-qubit 
$R_{ZZ}$ gates, especially in dense graphs. Compared to vanilla QAOA, DAPO uses only $66\%$  of 
$R_{ZZ}$ gates at the same depth while delivering better results, demonstrating its potential for efficient combinatorial optimization in the NISQ era.

}
\begin{description}
\item[Usage]
Secondary publications and information retrieval purposes.
\item[Structure]
You may use the \texttt{description} environment to structure your abstract;
use the optional argument of the \verb+\item+ command to give the category of each item. 
\end{description}
\end{abstract}

\maketitle


\section{\label{sec:level1}Introduction}

Solving combinatorial optimization problems on classical computers faces multiple challenges, such as exponentially growing search space and traps for local optimal solutions. These factors constrain the effectiveness of classical algorithms in addressing large-scale problems, often rendering it challenging to identify optimal solutions.
However, with the continuous development of quantum computing technology, quantum algorithms have achieved exponential acceleration in some cases by exploiting quantum parallelism. This trend has prompted many researchers to explore the potential of quantum computers in addressing these challenges.

Among these, the Quantum Approximate Optimization Algorithm \cite{farhi2014quantum} is a promising solution that has attracted significant attention due to its integration of quantum parallelism and the flexibility of classical optimization algorithms. This approach has opened new avenues for addressing combinatorial optimization problems \cite{ruan2020quantum,bravyi2022hybrid}. However, for large-scale or complex problems, enhancing the approximate solutions with QAOA often requires additional layers and increased circuit depth. In the NISQ era, this can be impractical due to limitations in decoherence times and gate error rates, which often lead to less effective solutions. Furthermore, increasing the number of quantum gates and circuit depth introduces more optimization parameters, which slows convergence and often yields suboptimal results.

To address these issues, researchers have proposed a series of optimization strategies to accelerate the optimization process and improve the approximation ratio. These strategies mainly focus on parameter optimization and circuit design improvements. In terms of parameter optimization, strategies such as selecting a good initial solution \cite{jain2022graph,lee2021parameters,sack2021quantum,shaydulin2019multistart,zhou2020quantum}, employing suitable optimizers \cite{acampora2023genetic,alam2020accelerating,fernandez2022study,lotshaw2021empirical,pellow2021comparison,yao2020policy,bonet2021performance,khairy2020learning}, incremental learning\cite{Gao2025} and using parameter symmetries \cite{zhou2020quantum,akshay2021parameter,shaydulin2021classical,shi2022multiangle} and transfer rules \cite{galda2021transferability,shaydulin2019multistart,shaydulin2021qaoakit} have effectively enhanced the efficiency and robustness of the algorithm. In terms of circuit design, researchers have proposed various improvements focusing on the two core components of QAOA: the mixer operator and the phase operator.

 Significant progress has been made in improving mixer operators, which has accelerated convergence and enabled QAOA to achieve optimal solutions with fewer layers, effectively reducing circuit depth. Notable advancements include, assigning independent parameters to each subterm improves the expressiveness of the operator by setting independent parameters for each mixer subterm \cite{herrman2022multi}. The introduction of adaptive bias fields incorporates dynamically adjusted bias terms into the Hamiltonian of the mixer, allowing the optimization process to converge more quickly to the optimal solution \cite{PhysRevResearch.4.023249}. An approach called Adaptive Derivative Assembled Problem Tailored-QAOA (ADAPT-QAOA) improves the algorithm's flexibility and performance by adaptively selecting mixer operators at each layer based on the maximum gradient principle \cite{zhu2022adaptive}. Furthermore, feedback-based parameter adjustment directly utilizes the measurement results of the quantum states to optimize the mixer parameters, eliminating the dependency on classical optimizers \cite{PhysRevLett.129.250502}. 
 
 However, despite these advancements, the scale and complexity of problems often cause phase operators to introduce a large number of CNOT gates, which become a primary factor in the increase of circuit depth. Therefore, simplifying phase operators is also a key factor. Research has also focused on optimizing phase operators.  Chandarana et al. combined QAOA with the concept of an antidiabatic drive, introducing an anti-diabatic term next to the original phase layer and mixer layer, thereby speeding up the algorithm's convergence and reducing circuit depth \cite{chandarana2022digitized}. The recursive method reduces problem variables layer by layer, shrinking the problem size and reducing computational complexity \cite{bravyi2020obstacles}. Additionally, in the ADAPT-QAOA, many terms in the vector of the optimal parameter tend to be zero, the operators corresponding to these terms do not affect the solution. Therefore, Yanakiev et al. remove these unnecessary phase operators, which reduces the depth of the circuit \cite{yanakiev2024dynamic}. However, this method does not directly simplify the structure of the phase operators themselves. In contrast, graph sparsification \cite{liu2022quantum} and QuantumDropout \cite{wang2023quantum} methods simplify the problem Hamiltonian, providing structural optimization of the phase operators, which effectively reduces the computational burden. But, these methods depend on the sparsity of the graph or the quality of the initial solution. If the sparsity is inadequate or the initial solution is of low quality, the phase operators derived from these solutions may propagate and amplify errors in the QAOA circuit. This can cause the circuit to become trapped in a local optimum and hinder convergence. 
Therefore, if the phase operators can be dynamically adjusted during the circuit construction process, it will help overcome these limitations and open up new possibilities for achieving more efficient convergence.

In this paper, we present a Dynamic Adaptive Phase Operator (DAPO) algorithm, which iteratively constructs phase operators layer by layer. The DAPO framework for building QAOA quantum circuits follows a similar structure to the vanilla QAOA, but with a key distinction: the phase operator for each subsequent layer is dynamically generated based on the output of the previous layer, rather than being fixed from the outset. By utilizing the solutions generated by QAOA itself to simplify the problem Hamiltonian at each layer, the algorithm better captures the problem's structural properties, progressively driving the solution closer to the optimal target. 
In addition, we use the neighborhood search algorithm to optimize the phase operator continuously, mitigating the risk of local optima while simultaneously accelerating the convergence of the ansatz to the optimal solution of the problem. To verify the effectiveness of the DAPO algorithm, we conducted experiments on the MaxCut problem and the NAE3SAT problem with different scales.
The experimental results show that the DAPO algorithm can not only achieve a higher approximation ratio but also reduces the number of $R_{ZZ}$ in the phase operator from $|E|$ (the number of edges of the original problem) to $|e|$ (the number of edges of the optimal solution). Its advantage is more prominent in dense graphs with a relatively small
optimal cut value.


Our paper is organized as follows. In Section \ref{preliminaries}, we introduce the principles of MaxCut problem and QAOA. In Section \ref{sec.DAPO}, we present the proposed DAPO algorithm and introduce the basis and technology of algorithm implementation in detail. Section \ref{experiment} evaluates our DAPO algorithm's approximate ratio and $R_{ZZ}$ number savings on MaxCut and NAE3SAT problems.
Finally, we provide conclusions in Section \ref{conclusion}.

\begin{figure}[h]     
\centering  
\includegraphics[width=8cm]{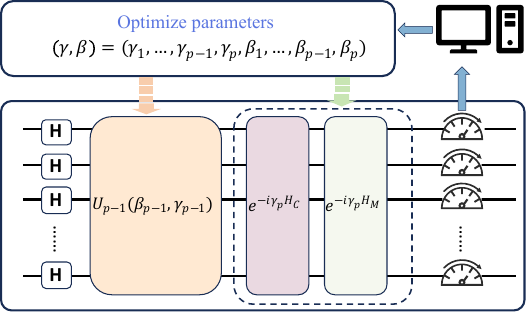}
\caption{In the $p$-layer schematic of vanilla QAOA, the algorithm constructs its circuit by alternating layers of problem Hamiltonian and mixer Hamiltonian parameterized quantum gates to maximize the expected value of the objective function, thereby achieving optimal solutions to optimization problems on quantum computers.} \label{vanilla-qaoa}
\end{figure}

\section{Preliminaries}\label{preliminaries}



In this section, we briefly introduce the principles of MaxCut. We then review the QAOA framework and the construction of the phase operator and mixer operator.

\subsection{Principles of MaxCut}

This paper expands on the MaxCut problem, a well-known combinatorial optimization problem that aims to decompose a set of vertices $V$ into two disjoint subsets to maximize the number of edges between the two subsets. The sets of edges with two endpoints in different subsets are \textit{cut}, and the one with the maximum size is \textit{optimal cut}.
 The Quadratic Unconstrained Binary Optimization (QUBO) model of the MaxCut problem can be expressed as:

\begin{equation}
   \text{Maximize} \quad C(x)=\sum_{(i,j)\in E}(x_i+x_j-2x_i x_j)
\end{equation}
where $E$ is the set of edges corresponding to the MaxCut problem, and $x=\{x_1,x_2,...,x_n\}\in \{0,1\}^{n}$.

The quantum Ising model and the QUBO model are closely related because they both involve the optimization of binary variables. The spin variable $Z_i=\pm 1$ in the Ising model and the binary variable $x_i\in \{0,1\}$ in the QUBO model can be converted to each other by a simple transformation $x_i =\frac{I+Z_i}{2}$.

\begin{equation}
    H_C = \frac{1}{2}\sum_{(i,j)\in E}(I-Z_i Z_j)
\end{equation}
where $Z_i$ is the Pauli $Z$ operator acting on the $i$-th qubit. Therefore, with this transformation, the QUBO problem can be solved by finding the ground state of the Hamiltonian corresponding to the Ising model. It also makes it possible to solve QUBO model using the quantum algorithm.

\subsection{Composition of QAOA}
The Hamiltonian $H_C$ corresponding to the Ising model is used to construct the phase operator of the vanilla QAOA. The vanilla QAOA is a quantum-classical hybrid algorithm whose quantum circuit is composed of phase operators and mixer operators. By optimizing the parameters in the parametric quantum gate with the classical optimizer, the approximate optimal solution of the problem can be obtained, as shown in Fig. \ref{vanilla-qaoa}. 


\textbf{Algorithm}: The vanilla QAOA is as follows.
\begin{itemize} 
\item[1] Initialize the quantum state: First, the qubits are initialized to a uniform superposition state.
\begin{equation}
    |+\rangle ^{\otimes n}=H^{\otimes n}|0\rangle^{\otimes n}
\end{equation}
\item[2] Build problem Hamiltonian: Translate the QUBO model of optimization problem into corresponding problem Hamiltonian $H_C$, where the ground state energy of $H_C$ corresponds to the minimum value of the problem's cost function.
\item[3] Building mixer Hamiltonian: Introducing mixer Hamiltonian evolves the quantum state from a manageable initial state to the ground state of $H_C$. Usually in a simple form, $H_M=\sum_i X_i$, where $X_i$ is the Pauli operator $X$ acting on the $i$-th qubit.
\item[4] Apply evolutionary operators: Combine the problem Hamiltonian and the mixer Hamiltonian to construct the evolution operators of the quantum state. 
\begin{equation}
    U(\beta,\gamma)=e^{-i\beta H_M}e^{-i\gamma H_C}
\end{equation}

Therefore, the evolution process of the  $p$-layer QAOA is:
\begin{equation}
    |\psi_p\rangle =U_p\cdots U_1|+\rangle^{\otimes n}
\end{equation}
\item[5] Optimization parameters: The key to QAOA is to optimize the parameters $\beta,\gamma=(\beta_p\gamma_p... \beta_1 \gamma_1)$, making the process of evolution as close as possible to the ground state. This is often achieved through classical optimization algorithms, such as gradient descent, simulated annealing and so on.
\end{itemize}

Repeat the adjustment parameters and evolution process, calculating the expectation value,
\begin{equation}
    F_p(\beta^*,\gamma^*) = \langle \psi_p(\beta^*,\gamma^*)|H_C|\psi_p(\beta^*,\gamma^*)\rangle
\end{equation}
until a certain stopping condition is met, such as reaching a maximum number of iterations or converging to a satisfactory accuracy.

\begin{figure}[h]     
    \centering     
    \includegraphics[width=8.5cm]{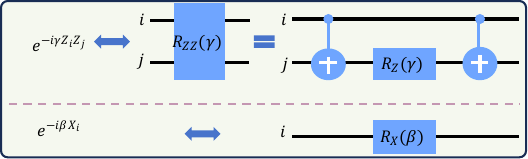}    
    \caption{Circuit for phase operator and mixer operator. Decomposition of the circuit for $e^{-i\gamma Z_iZ_j}=R_{zz}(\gamma)$ (upper) and the implementation of the circuit for $e^{-i\beta X}=R_X(\beta)$ (lower).}     
    \label{RZZ-RX} 
\end{figure}

As shown in Fig. \ref{RZZ-RX}, the mixer operator $e^{-i\beta X}$ corresponds to $R_X$ and the phase operator is composed of the problem Hamiltonian, expressed as $e^{-i\alpha H_C}$, where each $e^{-i\alpha Z_i Z_j}$ corresponds to the $R_{ZZ}$ operator operating on the $i$-th and $j$-th qubits. Since each edge in MaxCut problem leads to a $R_{ZZ}$ operator, the problem with high edge density will increase the number of quantum gates in the problem operator. Additionally, the number of parameterized gates increases with the layer count ($p$), leading to slower classical optimization, especially in noisy quantum circuits, where the growing number of gates further degrades solution quality.
 Therefore, whether a simple problem Hamiltonian can be used to solve a complex combinatorial optimization problem becomes the key to using the QAOA algorithm to solve large-scale problems in the NISQ era.

\begin{figure} [h]
\centering \includegraphics[width=8.5cm]  {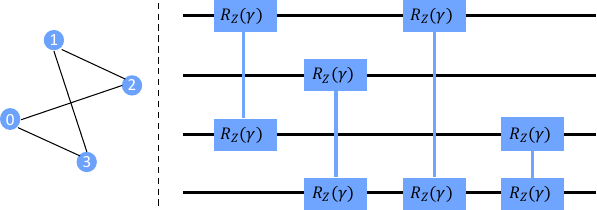}     \caption{Circuit implementation (right) of the problem hamiltonian $H_C$ corresponding to the MaxCut problem depicted in the graph (left).}     
\label{problem-Hamilton}
\end{figure}

Approximation ratio: When using approximation algorithms to solve combinatorial optimization problems, the approximation ratio is often used as an index to measure the quality of the solution.
\begin{equation}
    r=\frac{F(\beta^*,\gamma^*)}{C_{opt}}
\end{equation}
where $F(\beta^*,\gamma^*)$ is the value of the solution obtained by an approximation algorithm, and  $C_{opt}$ represents the optimal solution of the combinatorial optimization problem.  For a maximization problem like MaxCut, we want it to be as large as possible (ideally close to 1).

\section{Dynamic Adaptive Phase Operator algorithm}\label{sec.DAPO}

Considering that the gate density in the problem layer is the primary factor contributing to the increased circuit depth and slower solution speed for the MaxCut problem with a large number of edges, we focus specifically on optimizing the problem layer.
It is established that eliminating redundant edges absent from the cut set can simplify the Hamiltonian problem. For example, if we know the optimal cut edges, namely the solution, for a given MaxCut problem, we can construct the problem Hamiltonian based only on these edges, i.e., 
$ H_C = \frac{1}{2}\sum_{(i,j)\in \text{solution}}(I-Z_i Z_j)$.

\begin{figure}[h]      
\centering       
\includegraphics[width=8.5cm]
{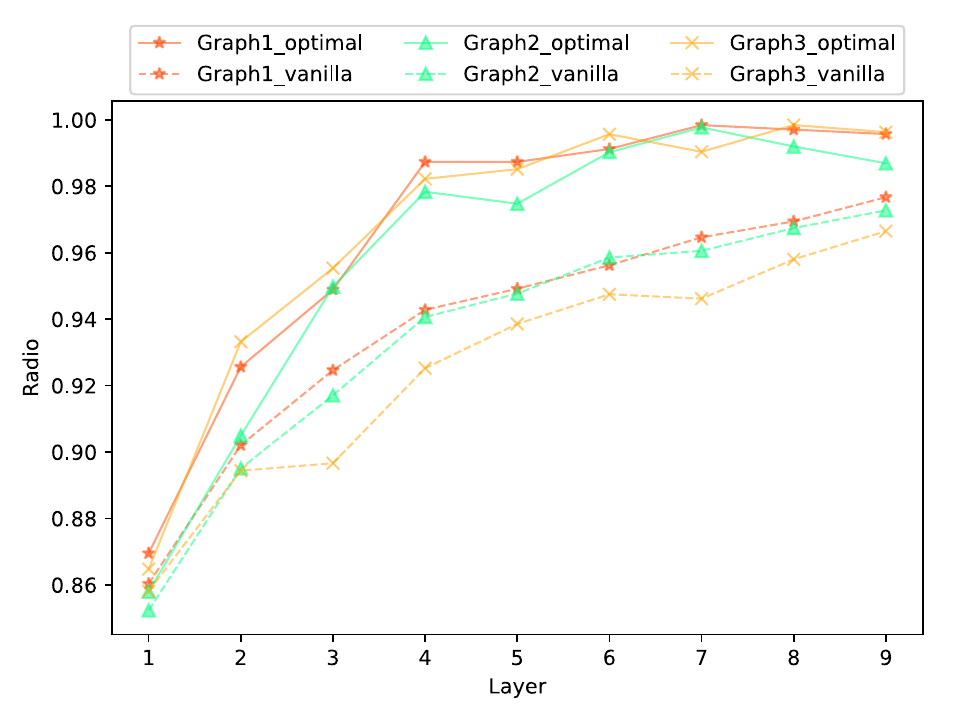}      
\caption{In three graphs each having 10 vertices with 30, 33, and 35 edges respectively, the maximum cut values are 20, 21, and 23. The figure illustrates the optimization results of Ansatz circuits constructed using optimal cut sets (blue) and vanilla QAOA circuits (red). From the results, it is evident that QAOA circuits constructed using optimal cut sets demonstrate better convergence.}  \label{fig5}  
\end{figure}

\begin{figure} [h]  
\centering     \includegraphics[width=8.5cm]{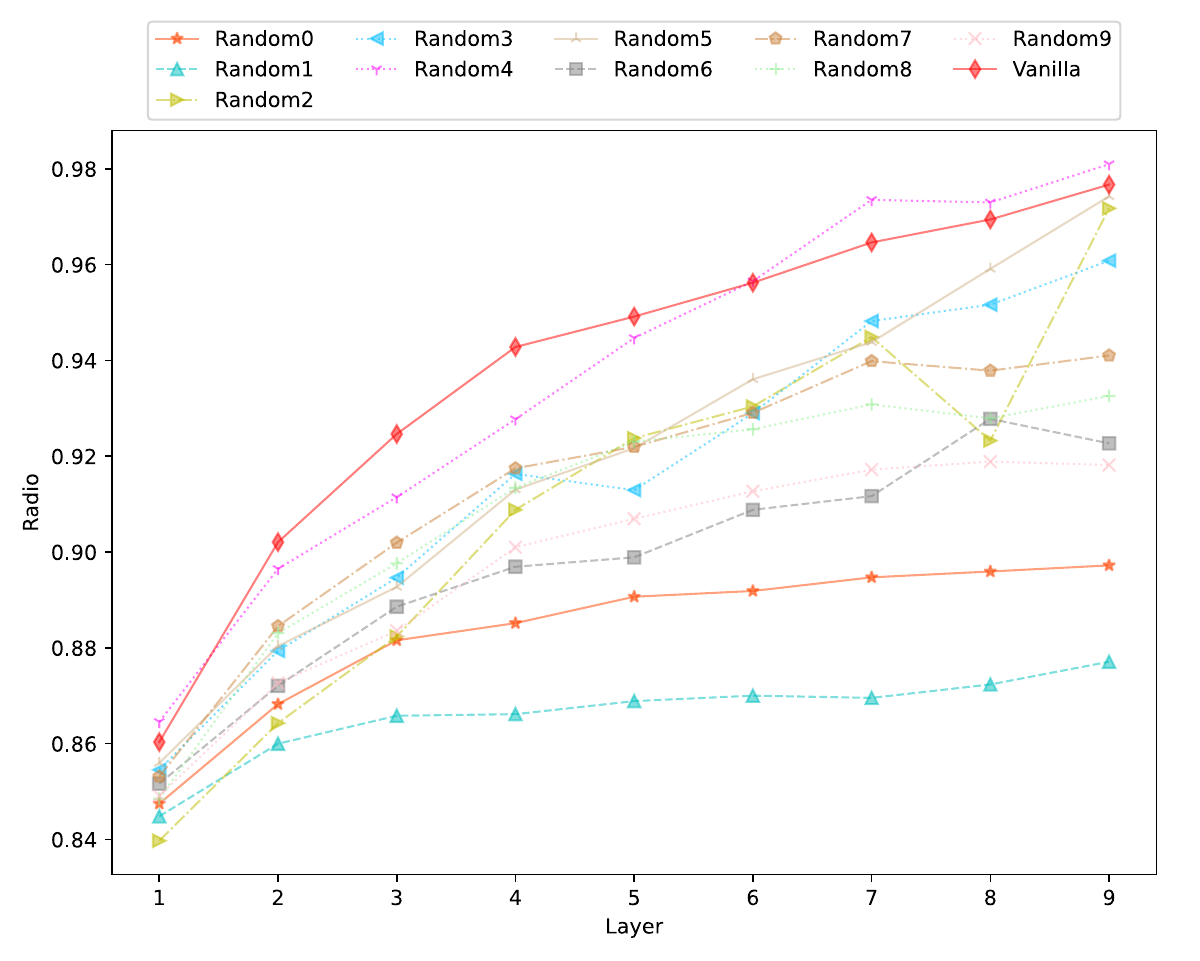}      
\caption{For a graph with 10 vertices and 30 edges, with a maximum cut value of 20, 10 sets of edges matching the number in the maximum cut value were randomly selected from the edge set $E$ to construct the Ansatz phase operator. The results indicate that Ansatz constructed from randomly generated edges generally approximates less closely compared to the vanilla QAOA circuit.}  \label{fig6} 
\end{figure}

Intuitively, using the cut solution to construct the problem Hamiltonian not only reduces circuit size, especially for large graphs but also improves convergence speed. As shown in Fig. \ref{fig5}, when the edge set corresponding to the optimal solution is used as the phase operator, the approximation ratio for solving the MaxCut problem is at least as good as, and sometimes better than, the vanilla QAOA.  However, the challenge lies in the fact that we often do not know in advance which edges are redundant or which will appear in the solution, making the task of simplifying the problem Hamiltonian difficult.
For sparse graphs or graphs with clear structure and fewer edges, random sparsification of the original Hamiltonian can work well since the simplified problem will not deviate much from the optimal cut. In such cases, the sparse Hamiltonian can capture the essential characteristics of the problem, leading to a reasonably good solution. However, for graphs with many edges, more targeted or structured, random sparsification tends to perform poorly, as illustrated in Fig. \ref{fig6}, where random sparse phase operators result in a worse approximation ratio than the vanilla QAOA. Thus more targeted or structured sparsification methods may be required to achieve good results. Other methods for simplifying the Hamiltonian, such as graph sparsification \cite{liu2022quantum}, and QuantumDropout \cite{wang2023quantum}, aim to optimize the structure of the phase operators. These methods reduce computational complexity but depend on the sparsity of the graph or the quality of the initial solution. If the sparsity is insufficient or the initial solution is poor, errors may propagate, amplifying inaccuracies in the QAOA circuit.

To address this, we propose a DAPO algorithm, which dynamically constructs phase operators incrementally. 


\begin{figure*}    
\centering     \includegraphics[width=6.0in]{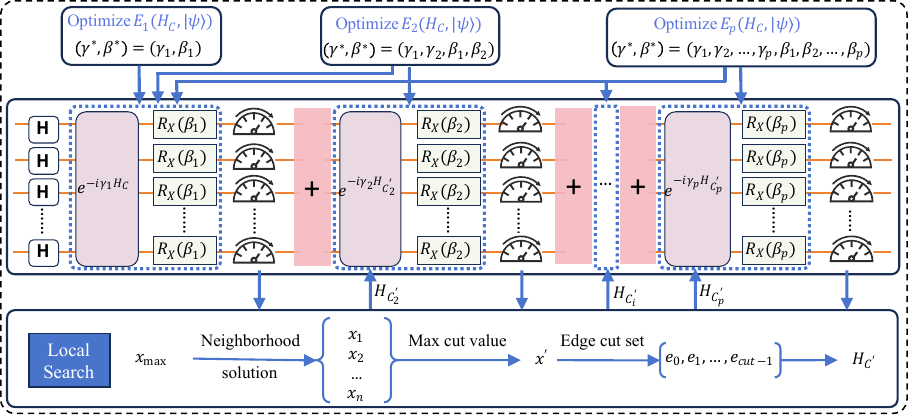}     
\caption{The iterative process of the DAPO algorithm for $p$ iterations is as follows: The system is first initialized to the maximum uniform superposition state. For $p=1$, the Ansatz circuit is constructed using the Hamiltonian $H_C$ related to the cost function. When $p>1$, the Ansatz circuit is iteratively built using the solution $x_{max}$ with the largest amplitude from the previous layer's circuit. And the mixer Hamiltonian being the $\sum_n{R_X}$ gates as in vanilla QAOA circuits. After obtaining $x_{max}$, the neighborhood search algorithm is used to find the solution $x^{\prime}$ with the highest objective function value within the neighborhood of $x_{max}$. The Hamiltonian $H_C^{\prime}$ corresponding to the cut set of $x^{\prime}$ is then used as the phase layer for the next Ansatz layer, rather than using the cut set corresponding to the measurement result $x_{max}$.} \label{DAPO}
\end{figure*}

\subsection{The details of DAPO}
As mentioned above, the phase operator corresponding to the optimal solution can lead to faster convergence, effectively simplifying the problem Hamiltonian. Although the optimal solution is unknown at the start, QAOA allows us to obtain the optimal solution for each layer during the iterative process. By adjusting the variational parameters ($\gamma$ and $\beta$), QAOA gradually refines the quantum circuit, driving the system toward the optimal solution. Based on this, we propose a dynamic adaptive approach in which the edge set from the current layer's optimal solution is used to construct the Hamiltonian for the next layer, progressively bringing the solution closer to the target optimal solution through successive layers.

The algorithm framework is shown in Fig. \ref{DAPO}. At the start of the algorithm, a uniform superposition state $(|s\rangle)$ is used as the initial state, representing the entire search space of the problem. In addition, the mixer operator is unchanged by $H_M=\sum_{i} X_i$ throughout the algorithm. In the first layer $(p=1)$, the phase operator remains the same as in vanilla QAOA, corresponding to the original problem Hamiltonian $H_C$.
\begin{equation}
    |\psi_1(\gamma,\beta)\rangle=(e^{-i\beta_1H_M}e^{-i\gamma_1H_C})|s\rangle
\end{equation}
The initial parameters $\gamma$ and $\beta$ are set to 0.01, and the circuit is optimized using a classical optimizer.
\begin{equation}     |\psi_1(\gamma^*,\beta^*)\rangle=(e^{-i\beta_1^*H_M}e^{-i\gamma_1^*H_C})|s\rangle \end{equation}

We choose the solution  $x_{max}$  with the largest amplitude in $|\psi_1(\gamma^*,\beta^*)\rangle$  as the basis for constructing the sparse phase operator for the next layer. By obtaining  $H_{C^{\prime}}$    corresponding to $x_{max}$, the sparse phase operator for the next layer can be derived. As the number of layers $p$ increases, we dynamically adjust the sparse phase operator based on the results from the previous layer, thereby obtaining the quantum circuit for layer $p$.
\begin{equation}     |\psi_p(\gamma,\beta)\rangle=(e^{-i\beta_k H_M}e^{-i\gamma_kH_{C^{\prime}}})|s\rangle
\end{equation}

Additionally, drawing inspiration from QuantumDropout \cite{wang2023quantum}, we preserve the cost function of the original problem while constructing the sparse phase operator. Since the cost function captures the physical and combinatorial constraints of the problem, it must remain intact to ensure that the global optimal solution's accuracy and physical meaning are preserved. This ensures that the DAPO algorithm gradually converges to the optimal solution while minimizing the risk of converging to a local optimum. Finally, optimize the dynamically constructed quantum circuit to obtain the final quantum state that converges to the optimal solution.

The framework of the DAPO algorithm for constructing QAOA quantum circuits is similar to that of the vanilla QAOA. The key difference is to dynamically generate the next layer's sparse phase operator based on the previous layer's output, rather than the fixed initial one. By utilizing the solutions generated by QAOA itself to simplify the problem Hamiltonian at each layer, we can better capture the structural properties of the problem. Compared to random sparsification methods and fixed classical sparsification methods, the proposed layer-by-layer optimization approach offers significant advantages. It effectively avoids the biases introduced by the randomization process, ensuring that each layer’s phase operator more accurately reflects the actual solution to the problem, thus improving the overall optimization performance. This provides a more efficient and reliable approach for solving related problems.

However, there is still a problem with this processing. If the algorithm falls into a local optimal solution, this characteristic may continuously spread in the circuit, resulting in slower convergence of the circuit. At the same time, when the number of layers is small since the solution information is not obvious enough, the possibility of falling into the optimal solution is greater, which will lead to a worse convergence situation.
To address this issue, we introduce the concept of neighborhood search, as discussed in \ref{NS}. By searching for a better solution within the neighborhood of the optimal solution found in the previous layer, we refine the phase operator and move closer to the global optimal solution.

\subsection{Neighborhood Search}
\label{NS}
Neighborhood Search Algorithm is a kind of heuristic algorithm used to solve optimization problems. Its core idea is to find a better solution by searching the neighborhood of the current solution. The algorithm starts with an initial solution, generates a set of neighborhood solutions, and selects the better solution through the evaluation function. 

\begin{figure*}       
\centering     \includegraphics[width=6.5in]{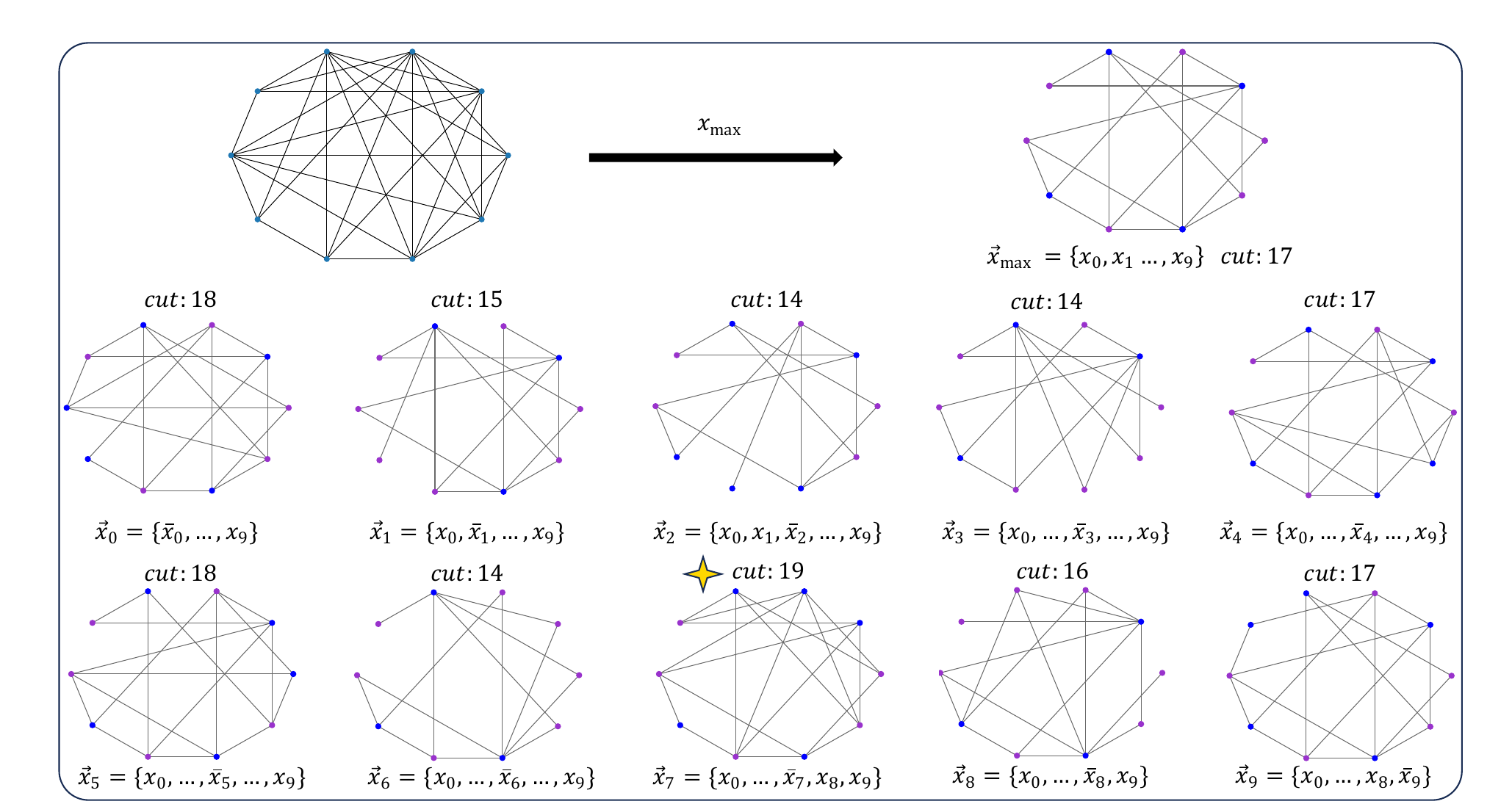}      
\caption{Neighborhood Search: Based on the measurement results $x_{max}$ from the previous layer, each bit $x_i$ in the bit string is sequentially flipped to $\bar{x_i}$. Using the updated string $x^{\prime}$, the vertex set of the graph is divided into two subsets, with edges connecting vertices in different subsets considered as cut edges. The cut value for each $x^{\prime}$ is calculated, and the one with the highest cut value is selected as the solution for constructing the cost layer in the next layer. The solution selected by the neighborhood search is marked with a pentagram in the graph.}  \label{fig7} 
\end{figure*}

One of the key problems in neighborhood search algorithms is how to choose the neighborhood. There are various ways to construct neighborhoods in neighborhood search algorithms, including distance-based construction, transformation-based construction, and combination-rule-based construction, and so on. In the DAPO algorithm, we adopt the transformation-based construction method. For instance, for a solution represented by a binary string, like $x = 10101$, we can generate neighboring solutions by randomly flipping one bit, and the resulting neighborhood solution set is $\{00101, 11101, 10001, 10111, 10100\}$. For a solution composed of $n$ bits, the neighborhood constructed in this way contains $n$ solutions.
During the construction of the Ansatz layer by layer, after obtaining the output result $x_{max}$ from the current layer, we search for a better solution within its neighborhood and construct the phase operator for the next layer based on the sparse Hamiltonian corresponding to the edge cut set of $x^{\prime}$, as illustrated in Fig. \ref{fig7}.


Let the current solution be $x$ and the objection function be $f$. The set of neighboring solutions is $N(x)$. We aim to find the solution within the neighborhood that maximizes the objective function value: 
\begin{equation}
    x^{\prime} = \mathop{\arg\max}\limits_{y\in N(x)} f(y)
\end{equation}
If $f(x^{\prime})>f(x)$, then update the current solution to $x^{\prime}$; otherwise, keep the current solution unchanged. In each iteration, the change in the objective function can be represented as: 
\begin{equation}
    \Delta f = f(x^{\prime})-f(x)
\end{equation}
When $\Delta f>0$, accept the new solution $x^{\prime}$ as the current solution. Otherwise, retain the original solution $x$. By following this process, the neighborhood search algorithm can effectively explore the solution space, aiming to find the global optimum or a solution close to the optimum.

\begin{algorithm}[H]
    \renewcommand{\algorithmicrequire}{\textbf{Input:}}
	\renewcommand{\algorithmicensure}{\textbf{Output:}}
	\caption{Dynamic-adaptive phase operator algorithm.}
    \label{alg}
    \begin{algorithmic}[1] 
        \REQUIRE  G(V, E).
	    \ENSURE Optimized circuit measurement results.
        
        \STATE Set initial state $|\psi_0\rangle =|+\rangle\dots |+\rangle ;$
        \STATE $p=0$;
        \STATE Random initialization parameters $\beta,\gamma$;
        
        \WHILE {Objective function not converge}
            \STATE $p = p + 1$;
            \IF {$p = = 1$}
                \STATE Apply the same phase operator as vanilla QAOA;
            \ELSE
                \STATE Apply sparse phase operator to Ansatz;
            \ENDIF
            \STATE Apply mixer operator as vanilla QAOA;
            \STATE Optimize parameters $\beta,\gamma$ for maximizing the $\langle \gamma,\beta|H|\gamma,\beta\rangle$;
            \STATE The solution $x$ with the largest amplitude is obtained by measuring the optimized circuit;
            \IF {$p == P$ or $|E_{p}^{*}-E_{p-1}^{*}|<\epsilon$}
                \STATE Objective function converged;
                \STATE The final optimized quantum circuit is obtained;
            \ENDIF
            \STATE Flipping every bit of $x$ yields $n$ neighborhood solutions to $x$;
            \STATE In $x$ and its $n$ neighborhood solutions, we find the solution $x_{max}$ with the highest cut value;
            \STATE The edge cut set of $x_{max}$ is obtained and the sparse phase operator is constructed for the next layer of Ansatz;
        \ENDWHILE
        \STATE Measure the optimized Ansatz circuit;

        \STATE \textbf{return} bit-strings.
    \end{algorithmic}

\label{alg01}   
\end{algorithm}

In the DAPO algorithm, incorporating a neighborhood search process ensures that the sparse phase operator approximates the phase operator
corresponding to the global optimal solution during the iteration process. Additionally, it effectively reduces circuit depth and the number of gates during layer-by-layer optimization, thereby improving the efficiency and convergence speed of the algorithm. The pseudo-code for the DAPO algorithm is shown in Algorithm \ref{alg01}.

\subsection{{$R_{ZZ}$ saved in DAPO}
}
\label{compare_cnot}
Here, we analyze the number of $R_{ZZ}$ gates saved by DAPO compared to the vanilla QAOA algorithm. To do so, we first clarify some basic concepts and theoretical bases. In quantum computing, each edge in the problem graph corresponds to an $R_{ZZ}$ gate in the phase operator. For a graph with $n$ vertices and $E$ edges, the set of cut edges corresponding to the optimal cut value is denoted as $e$. From graph theory, we know that $|e| \leq |E|$, meaning the number of cut edges cannot exceed the total number of edges in the graph.

The maximum cut value 
$|e|$  equals the total number of edges 
$|E|$  only in the case of a complete bipartite graph. Specifically, for a complete bipartite graph 
$K_{m,n}$, where $m$ and $n$ are the numbers of vertices in two disjoint vertex subsets respectively, the total number of edges is $mn$. When partitioning the graph by these subsets, the number of cut edges is also $mn$, hence $|e| = |E| = mn$.

However, in dense graphs, redundant edges that are not part of the optimal cut are often present. In the DAPO algorithm, a key feature is
constructing the phase operator layer by layer based on the optimal solution of the previous layer, while the neighborhood search is used to mitigate the local solution.
Let  $|e^{'}|$ represent the number of cut edges for a neighborhood solution.  Due to the nature of the neighborhood solution and its relationship with the optimal solution, we have  $|e^{'}| \leq |e|$. Combining this with $|e| \leq |E|$, it follows that $|e'| \leq |E|$.

This means the number of edges used to construct the phase operator in the DAPO algorithm is always less than or equal to the number of edges used in the vanilla QAOA algorithm. Particularly in dense graphs, where the optimal cut value $|e|$ is significantly smaller than 
$|E|$, the difference $|E| - |e|$ leads to a reduction in the number of $R_{ZZ}$ gates. 
For example, if  $|e| = k$ (where$ k << E$), the vanilla QAOA algorithm processes $|E|$ edges, while DAPO processes no more than $|e^{'}| \leq k$. The larger the difference between $|E|$ and $k$, the more $R_{ZZ}$ gates can be saved by DAPO.


\section{Experiment}\label{experiment}
In this section, we compare the approximation ratio of the proposed Dynamic Adaptive Phase Operator algorithm, the vanilla QAOA, and graph sparsification methods \cite{liu2022quantum} in constructing phase operators. Then, we compare the number of $R_{ZZ}$ used by different algorithms under the same layer number by experiments. Finally, the DAPO algorithm is verified on the NAE3SAT problem, which proves its generalizability.

We conducted experiments using the Qiskit quantum software package, with the L-BFGS-B optimizer chosen to optimize parameters in the circuit. This optimizer is a quasi-Newton method suitable for handling large-scale optimization problems with boundary constraints. In the experiments, we tested different graphs to evaluate the performance of the proposed DAPO algorithm under various conditions. Table \ref{tab:table4} provides detailed information on the number of vertices, edges, and the optimal cut value for each graph.

\begin{table}[b] 
\caption{\label{tab:table4}%
In the experiment, multiple graph instances were used to evaluate the performance of the proposed algorithm under different scales and complexities. The table lists detailed information for each graph, including the number of vertices, edges, and their corresponding optimal cut values.}
\begin{ruledtabular}
\begin{tabular}{cccc} 
Graph&Vertices& Edges& Optimal\\ 
\hline 
graph1&10&30&20\\
graph2&10& 33 & 22 \\
graph3&10& 35 & 23 \\
\end{tabular} 
\end{ruledtabular} 
\end{table}

\begin{figure*}        
\centering     
\includegraphics[width=7in]{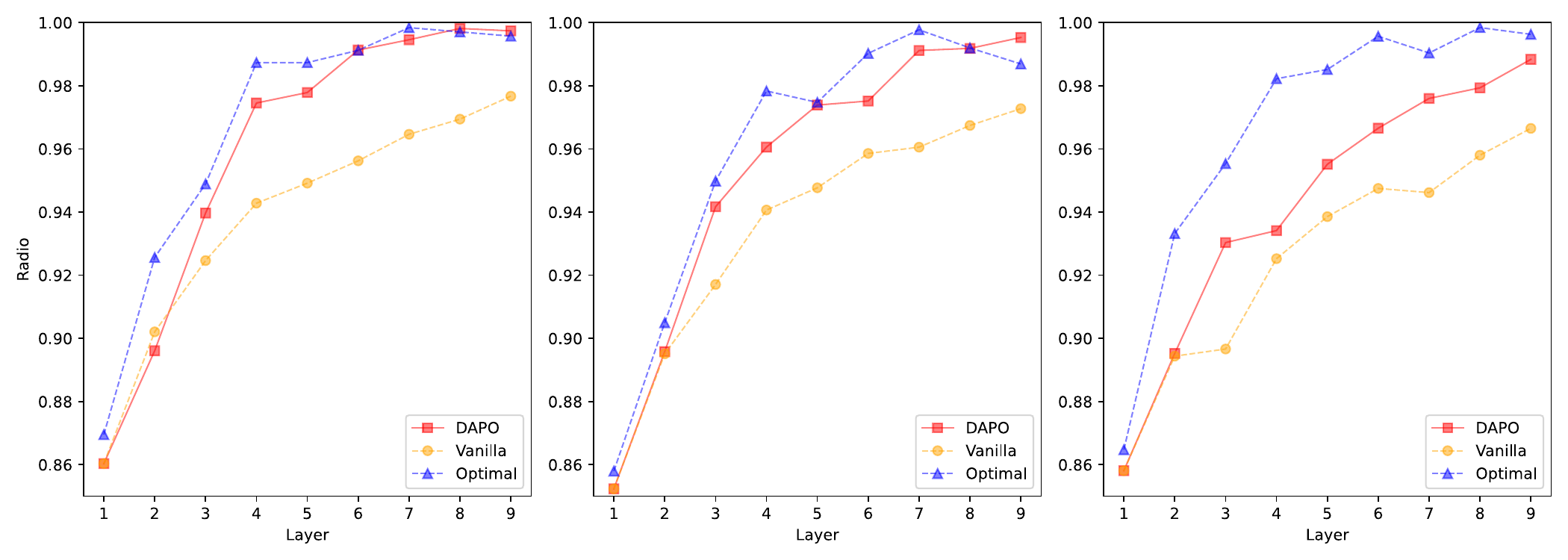}      
\caption{The approximation ratios of three QAOA algorithms on each layer of graph1, graph2, and graph3. Among these, the QAOA circuit constructed with the optimal solution (yellow) achieved the highest approximation ratios on all three graphs. The DAPO algorithm (red) obtained approximation ratios superior to those of the vanilla QAOA algorithm (blue) and exhibited faster convergence. As the number of layers increased, the approximation ratios achieved by the DAPO algorithm gradually approached those of the optimal circuit.}  
\label{fig8}  
\end{figure*}

\subsection{Approximation ratio of DAPO algorithm}
To highlight the advantages of the DAPO algorithm, we conducted comparative experiments between it and the standard QAOA algorithm, as well as the QAOA algorithm that constructs the phase operator using the global optimal solution. The experiments were carried out based on the problem graphs shown in Table \ref{tab:table4}, with a focus on examining the approximation ratio and convergence speed of each algorithm.

As shown in Fig. \ref{fig8}, with the maximum number of iterations of the optimizer set to 20,000, we compared the approximation ratios for solving the MaxCut problem using both the vanilla QAOA algorithm and our algorithm across different numbers of layers. The figure clearly illustrates that as the number of layers increases, the approximation ratios of both algorithms gradually improve. However, our algorithm demonstrates a more significant advantage, with its approximation ratio increasing at a faster rate than that of the vanilla QAOA algorithm. Overall, our algorithm consistently maintains a higher approximation ratio, especially with a greater number of layers. This result effectively proves that our algorithm converges faster and performs better compared to the vanilla QAOA algorithm. 

Meanwhile, in the DAPO algorithm, we continuously conduct refined optimization on the phase operator, aiming to make it approximate the phase operator corresponding to the global optimal solution as closely as possible. Consequently, it can be observed that as the optimization process progresses, the experimental performance exhibited by the DAPO algorithm is gradually approaching that of the QAOA algorithm which constructs the phase operator based on the global optimal solution, and the performance gap between the two is constantly narrowing.

\subsection{Comparison with graph sparse methods}

For more effective comparison and analysis, we employed the classic graph sparsification methods mentioned in \cite{liu2022quantum}. Graph sparsification aims to simplify the graph structure by reducing the number of edges while preserving the main structure and characteristics of the graph. This aligns with the goal of our sparse phase operators. Therefore, we used these methods to generate sparse graphs with the same number of edges as the maximum cut value and constructed phase operators based on these sparse graph edges to build the QAOA circuit.

\begin{figure*}     \centering     \includegraphics[width=7in]{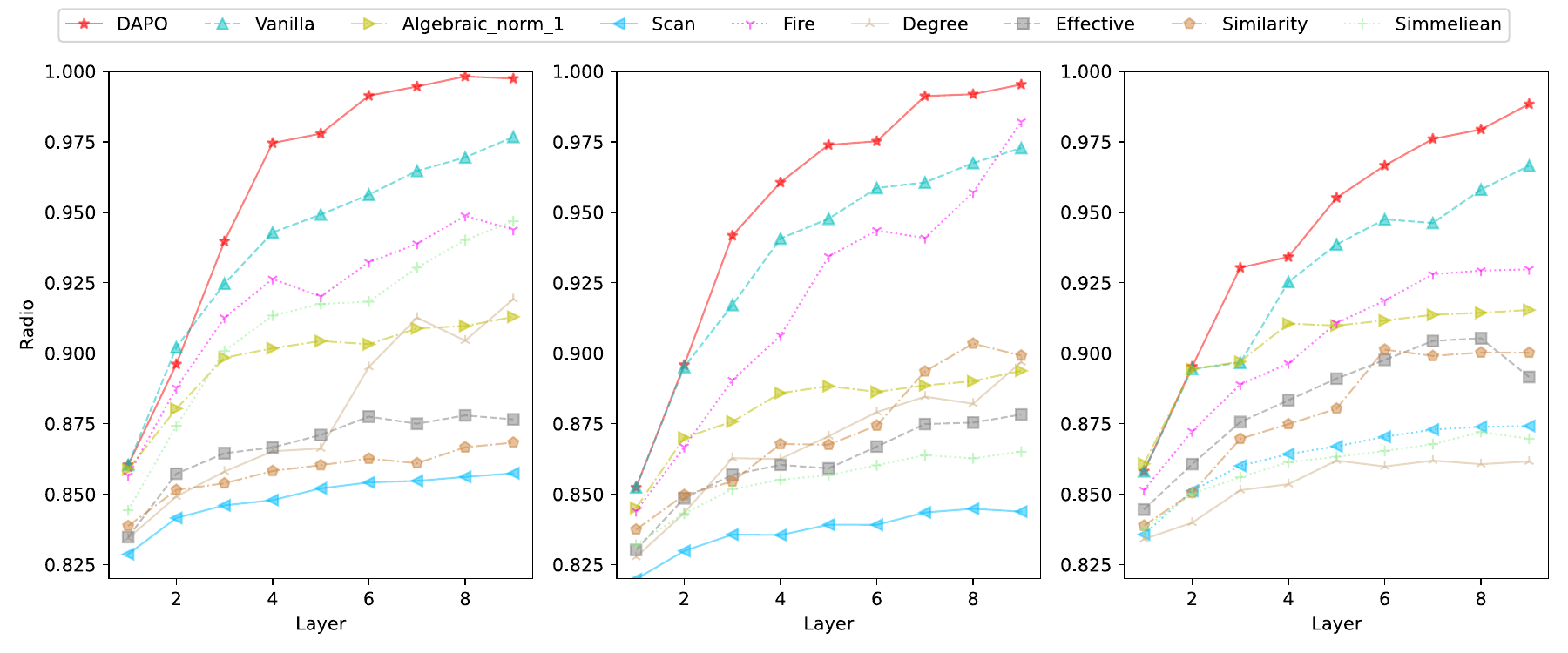}      
\caption{Approximate ratio of DAPO algorithm, vanilla QAOA, and QAOA Ansatz constructed by subgraphs corresponding to different sparsity methods on graphs graph1, graph2, and graph3. The results show that the DAPO algorithm achieved higher approximation ratios on all three graphs compared to the QAOA Ansatz constructed from subgraphs using other classical sparsification methods. However, it is not evident which specific classical method stands out as the best among all.} \label{fig9} \end{figure*}

From Fig. \ref{fig9}, it can be seen that none of the classical sparsification methods show a significant advantage over the others. Each method has its strengths and weaknesses depending on the graph structure and application scenario. However, our proposed DAPO algorithm demonstrates superior performance in several aspects. Firstly, the DAPO algorithm has better convergence, meaning it can achieve the desired results with fewer iterations. Secondly, while maintaining a high approximation ratio, the DAPO algorithm effectively reduces the number of edges in the graph. This reduction in edges decreases the number of CNOT, making our algorithm efficient in NISQ era.

\begin{figure*}       
\centering     
\includegraphics[width=7in]{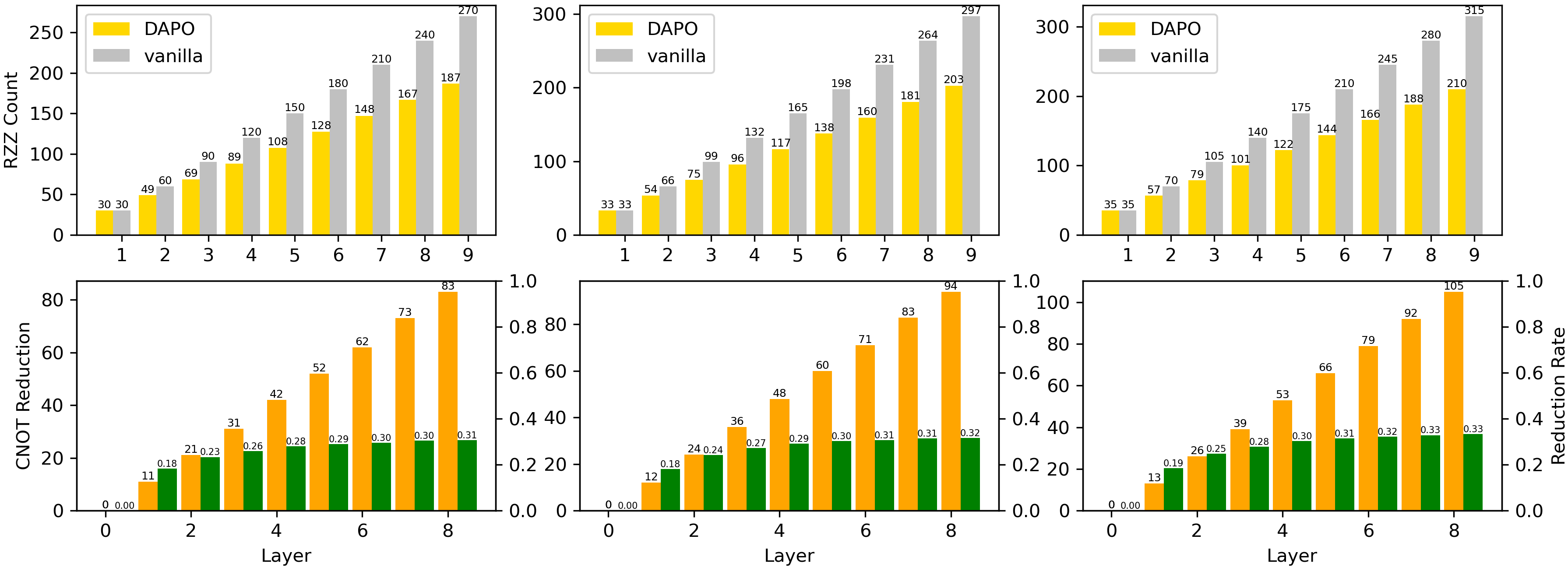}
\caption{The three figures in the above row show the number of $R_{ZZ}$ reduced by the DAPO algorithm at each layer of graph1, graph2, and graph3 compared with the vanilla QAOA algorithm. The three graphs in the following row show the reduction and proportion of the CNOT gate compared to the vanilla QAOA algorithm.}
\label{fig10}
\end{figure*}
\subsection{Comparison of the number of $R_{ZZ}$}

Now, through experiments, we have demonstrated the number of $R_{ZZ}$ saved by the DAPO algorithm compared with the vanilla QAOA, and further show the amount of CNOT saved.

As shown in Fig.\ref{fig10}, as the circuit depth increases, the number of $R_{ZZ}$ gates required by the DAPO algorithm increases linearly with the optimal cut value, while the number of $R_{ZZ}$ gates required by the vanilla QAOA increases linearly with the total number of edges in the problem instance. In other words, the DAPO algorithm requires fewer quantum gates compared to the vanilla QAOA, making it more efficient while maintaining accuracy. 
This also further corroborates the conclusion we presented in \ref{compare_cnot}, that is, the number of gates $|e^\prime|$ used in the phase operator of the DAPO algorithm is less than the number of edges $|E|$ of the original graph. 



\subsection{NAE3SAT}
Similarly, we validated our DAPO algorithm regarding NAE3SAT problems and conducted comprehensive experimental comparisons with Quantum Dropout and vanilla QAOA. Similar to the MaxCut problem, the NAE3SAT problem is also a combinatorial optimization problem. The NAE3SAT problem is a variant of the Boolean satisfiability problem (SAT). In the classic 3-SAT problem, given a Boolean formula that is in conjunctive normal form composed of several clauses, each clause contains exactly three literals, and the problem is to determine whether there exists a set of variable assignments that makes the entire formula true. However, in the NAE3SAT problem, it is required that for each clause, the assignments of the three literals cannot be all the same, that is, they cannot be all true or all false. 

The Hamiltonian of the NAE3SAT is as follows:
\begin{equation}
\begin{aligned}
    H_C=\sum_{i,j,k\in C}[(s_i+s_j+s_k)^2-1]/2 
    \\ = \sum_{i,j,k \in C}(s_is_j+s_js_k+s_ks_i)+|C|
\end{aligned}
\end{equation}
where $|C|$ is the number of clauses in $C$. When all clauses are true, $H_C$ takes the minimum value $0$, and the corresponding bit string is the solution of the NAE3SAT.

We verified the DAPO algorithm on hard NAE3SAT problems and five randomly generated NAE3SAT problem instances. Among them, the hard problems have 12 literals and 72 clauses, and the randomly generated instances have 10 clauses and 30 clauses. The difference between the easy and hard problem is the energy landscape. The global minimum is located in a large and smooth neighborhood for a simpler problem and a narrow region for a harder problem.

\begin{figure}[h]   
\centering     \includegraphics[width=8.5cm]{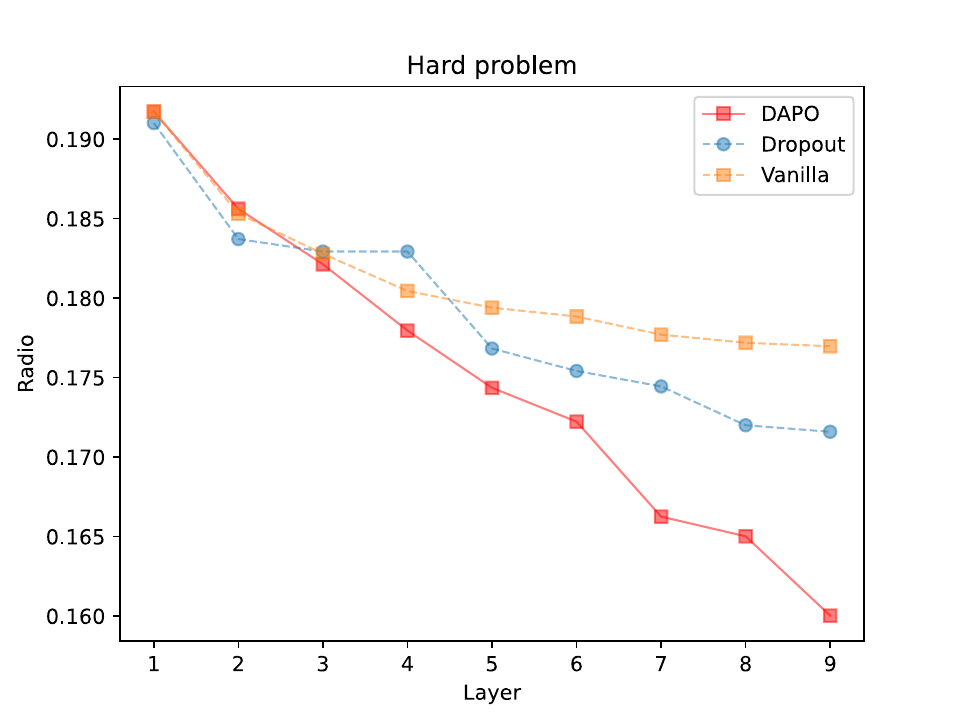}      
\caption{Approximate ratios of vanilla QAOA, DAPO, and Dropout on the hard NAE3Sat problem, where the number of layers p = 9. Among them, the dropout rate of the dropout algorithm $r$ = 0.5. The experiment results of Dropout are the best results of five random dropout.}  \label{fighardSAT} 
\end{figure}

\begin{figure}[h]       
\centering     \includegraphics[width=8.5cm]{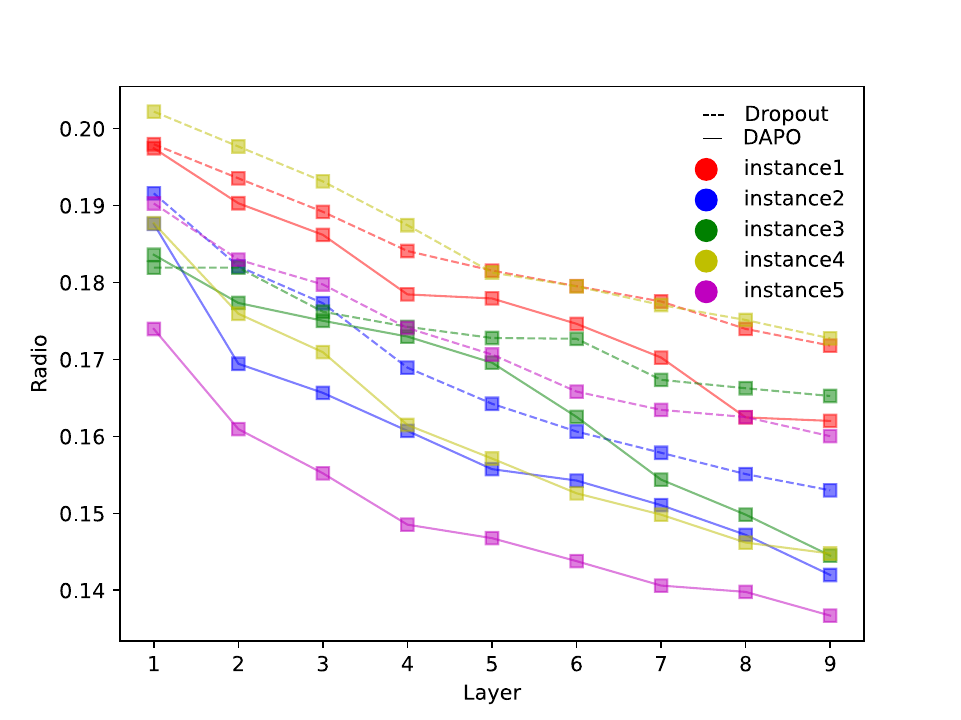}      
\caption{The approximate ratios of the DAPO algorithm and the Dropout method were evaluated on randomly generated NAE3SAT problem instances. In this experiment, the dropout rate for the Dropout algorithm was set to 0.5, with the results reflecting the best performance from five random runs. Additionally, each color represents the same problem instance, with dashed lines indicating the Dropout method and solid lines representing the DAPO method.}  \label{randomSat} 
\end{figure}

As shown in Fig. \ref{fighardSAT}, with a fixed number of layers, our DAPO algorithm is superior to the vanilla QAOA and dropout methods, achieving a lower energy difference. This enables it to approach the optimal solution more efficiently when facing challenging problems, leading to breakthroughs in problem-solving. For the randomly generated instances, Fig. \ref{randomSat} illustrates similarly outstanding performance. The results reveal an even better approximation ratio for these random examples, further highlighting the advantages of the DAPO algorithm in solving NAE3SAT problems. Our findings indicate that the DAPO algorithm excels not only in the MaxCut problem but also in the SAT context, achieving superior approximation ratios while utilizing fewer CNOT. These results underscore the potential of our approach to enhance the efficiency of quantum algorithms in tackling NP-complete problems, paving the way for future research in quantum optimization techniques.

\section{Conclusion}\label{conclusion}

In this paper, we introduced the Dynamic Adaptive Phase Operator (DAPO) algorithm, a novel quantum approach designed for combinatorial optimization problems. Building upon the principles of  QAOA, DAPO adaptively constructs sparse problem Hamiltonians by using the best solutions obtained at each iteration. Unlike traditional methods that rely on fixed or random sparsification strategies, DAPO dynamically adjusts the Hamiltonian at each layer based on the features of the current best solution, refined through neighborhood search. This dynamic adaptation allows for a progressive simplification of the Hamiltonian, bringing the solution closer to the global optimum with each successive layer.
As a result, we achieve an approximation of the ground state of the problem Hamiltonian using fewer $R_{ZZ}$ gates, with the number of $R_{ZZ}$ gates at each layer not exceeding the cut value.  

Additionally, by integrating the concept of neighborhood solutions, we selected better approximate solutions within the neighborhood of measurement results to form the edge cut set for the phase operator of the next layer. This ensures that the ground state solutions of the Hamiltonians corresponding to the Ansatz and the original problem are consistent.  

We tested our algorithm on three graphs with 10 vertices each, varying in density. Compared to the vanilla QAOA algorithm, our algorithm achieved higher approximation ratios using only 66\% of the $R_{ZZ}$ gates at the same depth. Furthermore, when compared with the QAOA algorithm applied to subgraphs sparsified by classical methods, our DAPO algorithm consistently yielded higher approximation ratios at the same depth.  In conclusion, our DAPO algorithm demonstrates the capability to achieve higher approximation ratios with fewer gates and reduced circuit depth. The significant reduction in the number of $R_{ZZ}$ gates underscores the potential of our algorithm to address the MaxCut problem more effectively in the NISQ era.
\begin{acknowledgments}
This research was supported by the National Nature Science Foundation of China (Grant No. 62101600,62301454), State Key Lab of Processors, Institute of Computing Technology, CAS under Grant No.CLQ202404, Fundamental Research Funds for the Central Universities (Grant No. SWU-KQ22049), and the Natural Science Foundation of Chongqing, China (Grant No. CSTB2023NSCQ-MSX0739).
\end{acknowledgments}










\nocite{*}

\bibliography{apssamp}

\end{document}